\documentclass[twocolumn,showpacs,aps,superscriptaddress]{revtex4}

\usepackage[dvips]{graphicx}
\usepackage{color}
\usepackage{amssymb}
\usepackage{amsmath}
\usepackage[latin1]{inputenc} 

\begin{document}
\title{Direct observation of domain wall structures in \\ curved permalloy wires containing an anti-notch}

\author{C.W. Sandweg}
\affiliation{Fachbereich Physik and Forschungsschwerpunkt MINAS,
Technische Universit\"at Kaiserslautern,
Erwin-Schr\"odinger-Stra\ss e 56, D-67663 Kaiserslautern, Germany}

\author{N. Wiese}
\affiliation{Department of Physics and Astronomy, University of
Glasgow, Glasgow G12 8QQ, United Kingdom}

\author{D. McGrouther}
\affiliation{Department of Physics and Astronomy, University of
Glasgow, Glasgow G12 8QQ, United Kingdom}

\author{S.J. Hermsdoerfer}
\affiliation{Fachbereich Physik and Forschungsschwerpunkt MINAS,
Technische Universit\"at Kaiserslautern,
Erwin-Schr\"odinger-Stra\ss e 56, D-67663 Kaiserslautern, Germany}

\author{H. Schultheiss}
\affiliation{Fachbereich Physik and Forschungsschwerpunkt MINAS,
Technische Universit\"at Kaiserslautern,
Erwin-Schr\"odinger-Stra\ss e 56, D-67663 Kaiserslautern, Germany}

\author{B. Leven}
\affiliation{Fachbereich Physik and Forschungsschwerpunkt MINAS,
Technische Universit\"at Kaiserslautern,
Erwin-Schr\"odinger-Stra\ss e 56, D-67663 Kaiserslautern, Germany}

\author{S. McVitie}
\affiliation{Department of Physics and Astronomy, University of
Glasgow, Glasgow G12 8QQ, United Kingdom}

\author{B. Hillebrands}
\affiliation{Fachbereich Physik and Forschungsschwerpunkt MINAS,
Technische Universit\"at Kaiserslautern,
Erwin-Schr\"odinger-Stra\ss e 56, D-67663 Kaiserslautern, Germany}

\author{J.N. Chapman}
\affiliation{Department of Physics and Astronomy, University of
Glasgow, Glasgow G12 8QQ, United Kingdom}

\date{\today}

\begin{abstract}
The formation and field response of head-to-head domain walls in curved permalloy wires, fabricated to contain a single anti-notch, have been investigated using Lorentz microscopy. High spatial resolution maps of the vector induction distribution in domain walls close to the anti-notch have been derived and compared with micromagnetic simulations. In wires of 10 nm thickness the walls are typically of a modified asymmetric transverse wall type. Their response to applied fields tangential to the wire at the anti-notch location was studied. The way the wall structure changes depends on whether the field moves the wall away from or further into the notch. Higher fields are needed and much more distorted wall structures are observed in the latter case, indicating that the anti-notch acts as an energy barrier for the domain wall.
\end{abstract}

\pacs{75.60.Ch 75.70.-i}

\maketitle

\section{Introduction}
The motion of domain walls (DWs) in small ferromagnetic elements has recently attracted much experimental and theoretical interest due to their application potential in magnetic logic \cite{Allwood} and data storage \cite{ParkinPatent} as well as for their fundamental physical properties \cite{Wieser}. DWs interact directly with electric currents and magnetic fields and their manipulation is of specific interest for the development of 'domain wall' electronics. Towards this end, a current-driven DW pendulum, working equivalently to a gravitational pendulum, has been implemented recently, providing information on the effective mass of a DW \cite{Saitoh}. However, it is also important to know the structure of the DW as a number of different possibilities have been identified in magnetic wires some 100s of nanometres in width and made from a soft magnetic material such as permalloy (Ni$_{81}$Fe$_{19}$). In particular, transverse domain walls (TDWs), asymmetric transverse domain walls (aTDWs) and vortex domain walls (VDWs) have been the subject of experimental and theoretical studies \cite{Nakatani,Handbook}. Moreover, for wires with geometries close to the phase boundaries for different wall types, transformation during the course of an experiment is not unusual
\cite{Junginger}. Hence direct observation to follow variation in the structure of a DW is frequently necessary.

\begin{figure}
\begin{center}
\includegraphics[width=8.6cm]{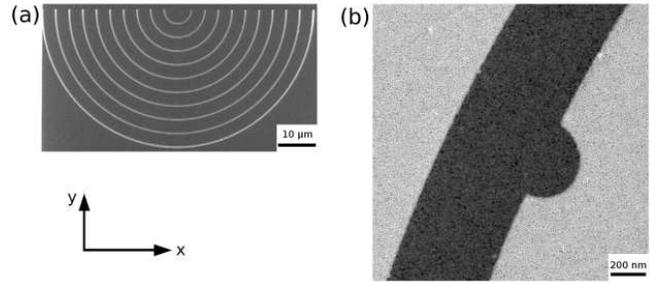}
\end{center}
\caption{\label{f:Fig1.eps}(a) SEM image of the fabricated permalloy wires, (b) low-magnification bright field TEM image showing an anti-notch.}
\end{figure}

Characterization of magnetic domains and the structure of the walls between them can be performed using various techniques including magneto-optical Kerr microscopy, magnetic force microscopy, X-ray magnetic circular dichroism and Lorentz microscopy \cite{Hubert}. In this paper, we report on Lorentz microscopy investigations of DW formation and manipulation in curved permalloy wires containing anti-notches or protuberances. Experimental results are compared with results obtained using micromagnetic simulation. The reason for fabricating notches or anti-notches in magnetic wires is primarily to provide locations where DWs can be pinned. For various applications it is important to know precisely where DWs are located and also to know how the geometry of the notch affects the pinning potential. The latter quantity depends not only on the notch itself but also on the DW structure. For this reason we have also investigated the DW structure as a function of applied field until it is pulled free of the anti-notch, leaving the magnetization on both sides aligned in a similar sense.

\section{Sample preparation}

The samples were produced using a combination of electron beam lithography and high vacuum evaporation. The required wire pattern was written in 120 nm thick PMMA (950K $4$\%) spun onto Si$_3$N$_4$ 'window' substrates of the kind used extensively in transmission electron microscopy (TEM) \cite{Khamsehpour2}. Such substrates contain one or more 100 $\mu$m square regions of unsupported Si$_3$N$_4$, $\sim$50 nm thick, and, as such, are essentially transparent to electrons. It is here that the wires to be studied were fabricated. After deposition of 10 nm of permalloy, lift-off of the PMMA in acetone left the desired pattern of wires.

Figure \ref{f:Fig1.eps} shows a scanning electron microscopy image of the sample. It comprises semi-circular wires with radii varying in steps of 5 $\mu$m from 5 $\mu$m  to 50 $\mu$m, the wire width being 500 nm. Such geometries are attractive since DWs can easily be created and subsequently moved by external magnetic fields \cite{Saitoh}. Wall creation is accomplished by letting the magnetization distribution relax following the application of a large field parallel to the symmetry axis (\emph{y}-direction). Movement of the DW along the wire then requires application of an orthogonal field (\emph{x}-direction). A low magnification bright field TEM image, fig. \ref{f:Fig1.eps}(b), shows that the wire is very well defined and also provides detail of the anti-notch located midway along the wire. The anti-notch was semicircular in shape with a radius of 250 nm.

\begin{figure}
\begin{center}
\includegraphics[width=8.6cm]{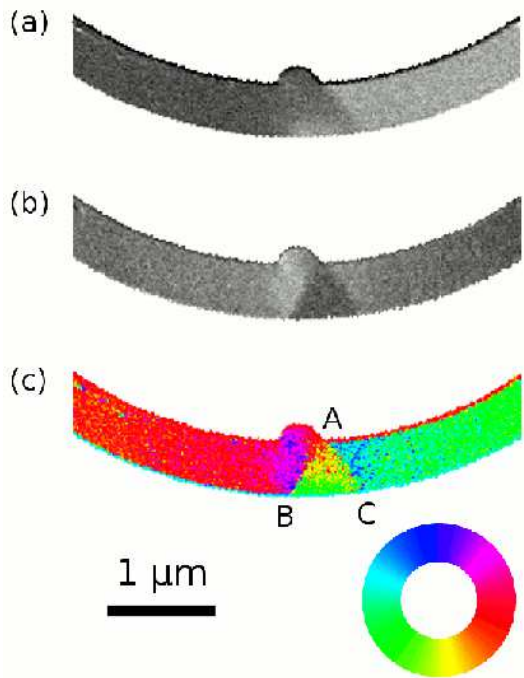}
\end{center}
\caption{\label{f:Fig2.eps}(color online) Remanent states of the curved wire with radius r = 5 $\mu$m after saturation in a field parallel to the  \emph{y}-axis. (a) and (b), DPC images of the  \emph{x}-  and  \emph{y}-components of magnetic induction, and (c), color induction map deduced from (a) and (b).}
\end{figure}

To determine the detailed form of the DWs generated, the differential phase contrast (DPC) mode of Lorentz microscopy was used \cite{Chapman1}. The TEM was a modified Philips CM20 equipped with (non-immersion) Lorentz lenses, thereby allowing magnetic imaging in a field-free environment with the standard objective lens switched off \cite{Chapman2}. In the experiments conducted here, the objective lens was weakly excited to provide a magnetic field suitable for moving the DW around. Control of the field to which the wires were subjected was achieved by tilting the TEM holder, thereby subjecting the specimen to a component of field in the plane of the wires. In the DPC imaging mode, pairs of images sensitive to orthogonal components of induction perpendicular to the electron trajectory are formed. The images are derived from the currents falling on opposite sectors of a quadrant detector and, as all the information is recorded simultaneously, are in perfect registration. Analysis of the image pairs then yields a quantitative vector map of averaged induction, vividly showing the detailed magnetic structure of the DWs under investigation. The resolution of the resulting map is determined principally by the probe diameter which in this instance was  $\approx$25 nm.

For comparison purposes, micromagnetic simulations were undertaken using OOMMF \cite{Donahue}. Parameters used were standard for permalloy (M$_{s}$ = 860 emu/cm$^{3}$, A = 1.3x10$^{-6}$ erg/cm and K = 0) whilst the cell size was 5 nm x 5 nm x 10 nm. The cell dimensions in the plane of the film were comparable with the material exchange length whilst that in the third dimension was set equal to the film thickness.

\section{Results}

\begin{figure}
\begin{center}
\includegraphics[width=7cm]{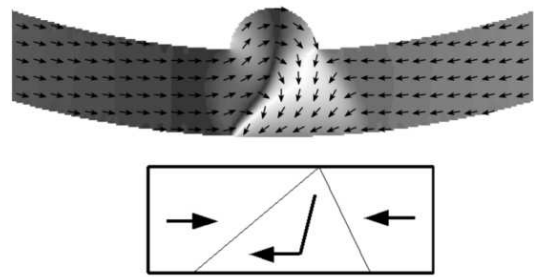}
\end{center}
\caption{\label{f:Fig3.eps}(color online) Micromagnetic simulation of the magnetization distribution in the vicinity of the anti-notch for comparison with fig. 2. A simplified schematic is shown below.}
\end{figure}

Figures \ref{f:Fig2.eps}(a) and (b) show respectively the  B$_x$ and B$_y$ components of induction of a DW at a notch whilst fig. \ref{f:Fig2.eps}(c) shows the calculated vector induction distribution displayed as a color map. The field used to create the DW was $\approx$ 3500 Oe and it was formed slightly away from the notch as alignment of the field with the \emph{y}-direction was always subject to a small error. However, application of a small {\em propagation field in the \emph{x}-direction, H$_{prop}$,} was all that was required to move the wall to the location shown in the figures. The wire used for fig. \ref{f:Fig2.eps} was the one with the smallest radius of curvature. Figure \ref{f:Fig3.eps} shows the equivalent micromagnetic simulation. It is clear that there is good agreement between experiment and simulation in terms of the general form assumed by the DW, which is of the head-to-head kind. The apex of the wall appears to be pinned close to the right hand side of the anti-notch (at A) whilst below the anti-notch a complex induction distribution exists. Thus in reality the DW should be thought of more as a DW packet \cite{McGrouther}, extending as it does over dimensions rather greater than the anti-notch diameter at the lower edge of the wire. The overall geometry is essentially that of an aTDW which can be represented schematically by two DWs of unequal length as shown in the inset of fig. \ref{f:Fig3.eps}. Here, the principal difference from a standard aTDW is due to the influence of the non-uniform magnetisation distribution in the anti-notch, in which there is inevitably some circulation of flux. Despite this, as can be seen most clearly from the arrows in fig. \ref{f:Fig3.eps}, the circulation is never complete and no vortex can be discerned, as would be the case for a vortex domain wall (VDW).

\begin{figure}
\begin{center}
\includegraphics[width=8.6cm]{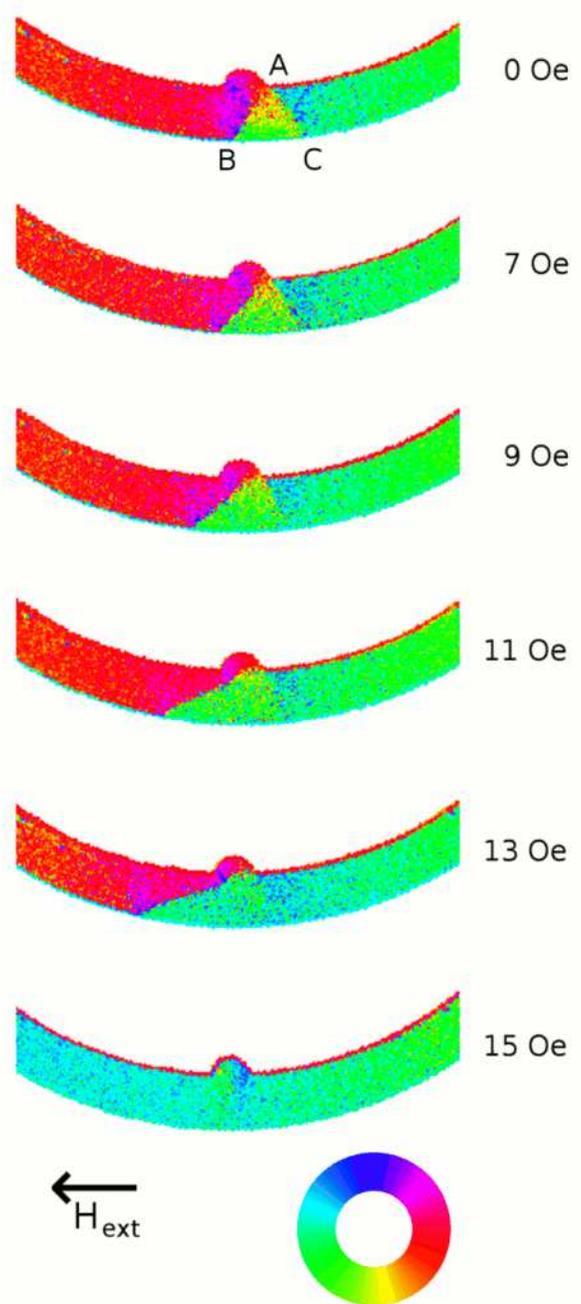}
\end{center}
\caption{\label{f:Fig4.eps}Color vector maps of the magnetic induction in a curved wire with radius r =  5 $\mu$m under application of increasing magnetic fields in the negative \emph{y}-direction.}
\end{figure}

To test the stability of the DW around the anti-notch position a small field was applied in the negative \emph{x}-direction. Application of a field in this direction should push the DW further into, and subsequently through, the anti-notch. The results are shown in fig. \ref{f:Fig4.eps}. At a field of $7$ Oe a modest extension ($\sim 25$ {\%}) of the DW width along the bottom edge of the wire can be seen. Under further field increase, the wall marked AB in the original DW packet became more pronounced and the end B moved progressively away from the anti-notch whilst the wall marked AC became increasingly indistinct as the wall angle decreased. At $13$ Oe, B was displaced by almost 1 $\mu$m from its original position and by $15$ Oe the DW was completely depinned leaving the wire essentially uniformly magnetized other than in the immediate vicinity of the anti-notch. Here there was still significant variation in the induction orientation, the magnetization close to the edge of the anti-notch trying to follow its contour, thereby reducing magnetostatic energy at the expense of a modest increase in exchange energy.

\begin{figure}
\begin{center}
\includegraphics[width=8.6cm]{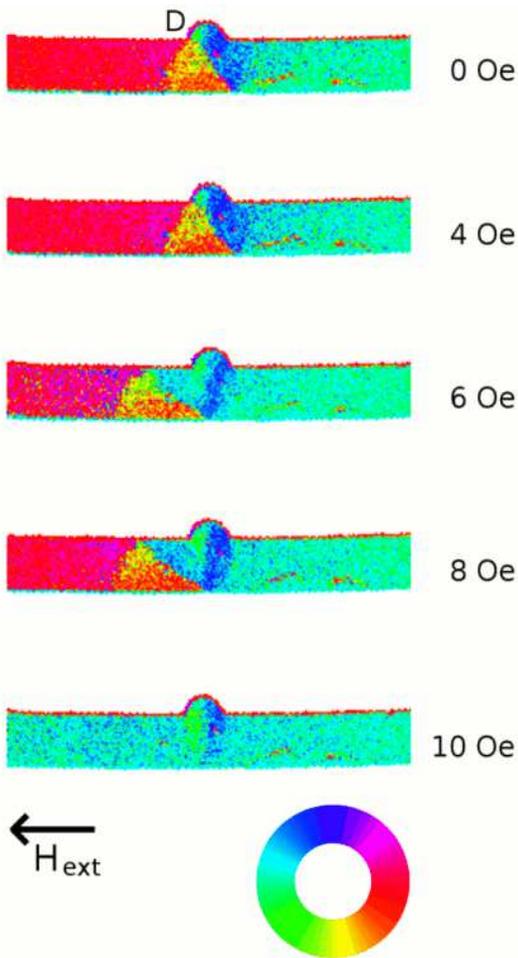}
\end{center}
\caption{\label{f:Fig5.eps}Color vector maps of the magnetic induction in a curved wire with radius r = 50 $\mu$m under application of increasing magnetic fields in the negative \emph{y}-direction.}
\end{figure}

Not dissimilar results were obtained when a DW was formed, in the same way as before, in the wire with the largest radius of curvature. Figure 5(a) shows the induction color map derived from a pair of DPC images when the \emph{y}-axis field was reduced to zero. Once again a head-to-head DW resulted although in this instance the apex was on the opposite side of the anti-notch at location D. This is likely to be the result of a very small change in field direction and, as such, is not important here. Again a field in the negative \emph{x}-direction was applied to the wire but now it should be noted that the effect of the field was to cause the wall to move away from, rather than into, the anti-notch. The results are shown in the remainder of fig. \ref{f:Fig5.eps}. For fields up to $4$ Oe there was no discernible change in the DW structure. However, by a field of $6$ Oe, both the DW location and overall width had changed. The wall apex moved $\sim 500$ nm from D whilst the width along the lower edge increased by $\sim 50$ {\%}. Further field increase had little effect until at $10$ Oe the DW moved completely from the field of view, leaving a quasi-uniformly magnetized wire.

To ascertain whether the behavior of the walls shown in figs. \ref{f:Fig4.eps} and \ref{f:Fig5.eps} was typical, the experiments were repeated many times with observations being made on the wires with radii 5, 40, 45 and 50 $\mu$m. Qualitatively this turned out to be the case. Moreover, in all cases it was found that the field required to pull an aTDW away from the anti-notch (depinning field, H$_{depin}$) was significantly lower than that required to push the aTDW through it (transmission field, H$_{trans}$). For the three wires with the largest radii the mean fields were found to be H$_{depin} = (4.0 \pm 1.0)$ Oe and H$_{trans} = (8.8 \pm 1.3)$ Oe. These were significantly larger than the propagation field necessary to move the aTDW towards the notch, that was found to be H$_{prop} = (1.2 \pm 0.4)$ Oe. The value of the transmission field is considerably smaller than the fields seen in fig. \ref{f:Fig4.eps} and further observations confirmed that there was a clear increase in this field with decreasing wire radius. A final observation relates to what happened to the DWs at fields just above those required to pull them away from the anti-notch, typically fields in the $4$ to $7$ Oe range. Here there was considerable variation in the distance that the DW moved, presumably due to the existence of local DW pinning sites. A good example of such local pinning was apparent in fig. \ref{f:Fig5.eps} where the DW remained at the same location some 500 nm away from the anti-notch until a field of $10$ Oe was applied. Study of structural TEM images revealed no obvious defect about the pinning site and certainly local pinning did not occur at the same distance from the anti-notch for other wires. Further work is ongoing to study such variations in behavior of nominally identical wires fabricated in a single batch.

\section{Discussion and Conclusion}

The formation and field response of head-to-head DWs in curved permalloy wires fabricated with a single anti-notch have been investigated using Lorentz-microscopy. Specifically, high spatial resolution color maps of the vector induction distribution in the vicinity of the anti-notch have been derived from pairs of DPC images. At remanence, the DW is in the form of an aTDW with modest magnetization circulation occurring in the anti-notch itself. Experimental results and micromagnetic simulations are in good agreement. Moreover, for wires with the radii of curvature studied here, the remanent DW structure was essentially independent of this parameter.

Detailed information on the way the DW responds to applied fields tangential to the wire at the anti-notch location was obtained by analyzing pairs of DPC images recorded under different fields. As recording times are typically 15 s Lorentz TEM is a very efficient technique for such in situ experimentation. Here we considered two cases, one where the effect of the field was to move the wall away from the anti-notch and the other where the wall had to be pushed through the anti-notch before a state of 'uniform' magnetization was realized. In the former case, fields $\approx4$ Oe were required to move the apex of the wall away from the anti-notch. For the latter case, where the wall had to be pushed through the anti-notch, rather higher fields were involved. This is not surprising as the apex of the wall appeared firmly anchored to the point where the edge of the anti-notch meets the uniform wire. The experimental results show very clearly that the shorter of the two DW segments making up the aTDW essentially disappears whilst the longer increases further in length and becomes closer to a 180$^{\circ}$ wall in character. As the specific wall energy increases with wall angle \cite{Hubert} significant Zeeman energy is presumably required to effect this change. Thus the field strength required to move a DW from the vicinity of an anti-notch depends very strongly on the side of the anti-notch where the apex of the wall is located and the direction of the applied field. 

The dependence of pinning and depinning fields on the notch geometry has been described previously with respect to the energy landscape a DW packet experiences in the vicinity of an artificial pinning site \cite{Petit2007}. Also the influence of triangular notches on TDWs was recently studied in detail by electron holography \cite{Klaeui2005,Backes2007}. Here, due to the reduced size of the DW package inside the notch, the total energy of the transverse wall can be significantly reduced so that the notch acts as a potential well. In contrast, in the case of the anti-notch presented above, the aTDW is pinned in front of the artificial pinning site, which indicates that the anti-notch represents an energy barrier which the DW has to overcome before it can continue to propagate. Furthermore, our experiments show that the depinning and transmission fields from the notch are larger than the propagation field necessary to move the DW along the wire towards the notch at the beginning of the field sequence. Therefore, the anti-notch acts like an energy barrier with small potential wells at both sides. It should be noted that the chosen geometry acts as a weak pinning potential in comparison to other shapes, as reflected by the comparably low fields for depinning and transmission \cite{Klaeui2005, Petit2007}. 

In summary, the complex behavior of DWs in magnetic wires with an anti-notch has been studied directly and in great detail using high resolution Lorentz microscopy. From the investigation the overall shape of the pinning potential has been determined.

\section{Acknowledgment}

The authors acknowledge the Nano+Bio Center of the Kaiserslautern University of Technology for technical support and Andreas Beck for sample preparation. Financial support by the European Commission within the EU-RTNs SPINSWITCH (MRTN-CT-2006-035327) and MULTIMAT (MRTN-CT-2004-505226) and by the DFG within the SPP1133 is gratefully acknowledged. The views expressed are solely those of the authors, and the other Contractors and/or the European Community cannot be held liable for any use that may be made of the information contained herein.

\end{document}